\documentclass[galaxies,review,accept,pdftex,oneauthor]{Definitions/mdpi} 
\usepackage{acronym}
\usepackage{xspace}
\usepackage{comment}
\firstpage{1} 
\pubvolume{1}
\issuenum{1}
\articlenumber{0}
\pubyear{2026}
\copyrightyear{2026}
\datereceived{4 June 2026} 
\daterevised{17 July 2026} %
\dateaccepted{24 July 2026} 
\datepublished{ } 
\Title{The first decade of gravitational-wave measurements of black hole spins}

\Author{Sylvia Biscoveanu $^{1}$*\orcidA{}}

\AuthorNames{Sylvia Biscoveanu}

\address{%
$^{1}$ \quad Department of Physics, Princeton University, Princeton, NJ 08544, USA; sbisco@princeton.edu}

\corres{Correspondence: sbisco@princeton.edu}

\abstract{A decade after the first direct detection of gravitational waves, the growing catalog of \textcolor{black}{hundreds of} confirmed events is revealing new insights into the spins of stellar-mass black holes. Spin measurements have long been heralded as a promising tracer of compact-object binary formation and evolution, as different formation channels predict \textcolor{black}{distinct} spin signatures on a population level. 
In this review, we summarize the astrophysics, phenomenology, and current measurements of black hole spins. We begin with an overview of the predictions for black hole spin magnitudes and orientations from leading formation channels---isolated binary evolution, dynamical formation in clusters, formation in AGN disks, and hierarchical triples. We then describe the imprint of spin effects on the gravitational waveform and the measurability of spin in individual events. Finally, we review current population-level constraints on spin magnitudes, orientations, and effective spin parameters, including correlations with mass and redshift, and discuss their astrophysical implications. We conclude by highlighting open questions and future prospects, emphasizing how improved detector sensitivity will enable increasingly precise spin measurements for both individual events and the binary black hole population as a whole.}


\begin{document}

\acrodef{O4}[O4]{fourth observing run}
\acrodef{BBH}[BBH]{binary black hole}
\acrodef{IID}[IID]{independently and identically distributed}
\acrodef{LVK}[LVK]{LIGO-Virgo-KAGRA}
\acrodef{PN}[PN]{post-Newtonian}
\acrodef{SOR}[SOR]{spin-orbit resonance}
\acrodef{RMR}[RMR]{reversed mass ratio}
\acrodef{SMR}[SMR]{standard mass ratio}
\acrodef{SNR}[SNR]{signal-to-noise ratio}
\acrodef{VM}[VM]{von Mises}
\acrodef{KDE}[KDE]{Kernel Denisty Estimate}
\acrodef{MC}[MC]{Monte Carlo}
\acrodef{MCMC}[MCMC]{Markov Chain Monte Carlo}
\acrodef{CDF}[CDF]{cumulative distribution function}
\acrodef{PPD}[PPD]{posterior predictive distribution}
\acrodef{GW}[GW]{gravitational-wave}
\acrodef{PSD}[PSD]{power spectral density}
\acrodef{CHE}[CHE]{chemically homogeneous evolution}
\acrodef{AGN}[AGN]{active galactic nuclei}
\acrodef{SMT}[SMT]{stable mass transfer}
\acrodef{2G}[2G]{second-generation}
\acrodef{NSBH}[NSBH]{neutron star-black hole}
\acrodef{FAR}[FAR]{false alarm rate}

\newcommand{\chieff}{\ensuremath{\chi_{\mathrm{eff}}}\xspace}
\newcommand{\OThreeAsymMaxP}{0.33}
\newcommand{\OThreeAsymMinus}{0.22}
\newcommand{\OThreeAsymPlus}{0.19}
\newcommand{\OThreeAsymUncert}{\ensuremath{\OThreeAsymMaxP_{-\OThreeAsymMinus}^{+\OThreeAsymPlus}}}
\newcommand{\OThreeHeavyMaxP}{0.95}
\newcommand{\OThreeHeavyMinus}{0.78}
\newcommand{\OThreeHeavyPlus}{0.05}
\newcommand{\OThreeHeavyUncert}{\ensuremath{\OThreeHeavyMaxP_{-\OThreeHeavyMinus}^{+\OThreeHeavyPlus}}}
\newcommand{\OFourHeavyMaxP}{0.97}
\newcommand{\OFourHeavyMinus}{0.29}
\newcommand{\OFourHeavyPlus}{0.03}
\newcommand{\OFourHeavyUncert}{\ensuremath{\OFourHeavyMaxP_{-\OFourHeavyMinus}^{+\OFourHeavyPlus}}}
\newcommand{\HierarchicalFirstMaxP}{0.78}
\newcommand{\HierarchicalFirstMinus}{0.09}
\newcommand{\HierarchicalFirstPlus}{0.08}
\newcommand{\HierarchicalFirstUncert}{\ensuremath{\HierarchicalFirstMaxP_{-\HierarchicalFirstMinus}^{+\HierarchicalFirstPlus}}}
\newcommand{\HierarchicalSecondMaxP}{0.62}
\newcommand{\HierarchicalSecondMinus}{0.34}
\newcommand{\HierarchicalSecondPlus}{0.37}
\newcommand{\HierarchicalSecondUncert}{\ensuremath{\HierarchicalSecondMaxP_{-\HierarchicalSecondMinus}^{+\HierarchicalSecondPlus}}}
\newcommand{\OFourHeavyMassMaxP}{242.96}
\newcommand{\OFourHeavyMassMinus}{29.18}
\newcommand{\OFourHeavyMassPlus}{23.77}
\newcommand{\OFourHeavyMassUncert}{\ensuremath{\OFourHeavyMassMaxP_{-\OFourHeavyMassMinus}^{+\OFourHeavyMassPlus}}}
\newcommand{\OThreeHeavyMassMaxP}{150.10}
\newcommand{\OThreeHeavyMassMinus}{18.93}
\newcommand{\OThreeHeavyMassPlus}{31.66}
\newcommand{\OThreeHeavyMassUncert}{\ensuremath{\OThreeHeavyMassMaxP_{-\OThreeHeavyMassMinus}^{+\OThreeHeavyMassPlus}}}
\newcommand{\OThreeAsymMassRatioMaxP}{0.28}
\newcommand{\OThreeAsymMassRatioMinus}{0.07}
\newcommand{\OThreeAsymMassRatioPlus}{0.20}
\newcommand{\OThreeAsymMassRatioUncert}{\ensuremath{\OThreeAsymMassRatioMaxP_{-\OThreeAsymMassRatioMinus}^{+\OThreeAsymMassRatioPlus}}}
\newcommand{\HierarchicalFirstMassRatioMaxP}{0.31}
\newcommand{\HierarchicalFirstMassRatioMinus}{0.09}
\newcommand{\HierarchicalFirstMassRatioPlus}{0.08}
\newcommand{\HierarchicalFirstMassRatioUncert}{\ensuremath{\HierarchicalFirstMassRatioMaxP_{-\HierarchicalFirstMassRatioMinus}^{+\HierarchicalFirstMassRatioPlus}}}
\newcommand{\HierarchicalSecondMassRatioMaxP}{0.42}
\newcommand{\HierarchicalSecondMassRatioMinus}{0.18}
\newcommand{\HierarchicalSecondMassRatioPlus}{0.28}
\newcommand{\HierarchicalSecondMassRatioUncert}{\ensuremath{\HierarchicalSecondMassRatioMaxP_{-\HierarchicalSecondMassRatioMinus}^{+\HierarchicalSecondMassRatioPlus}}}

\section{Introduction}
A decade after the first direct detection of gravitational waves, the catalog of sources detected by the \ac{LVK} collaboration has grown to include hundreds of \ac{BBH} mergers, two binary neutron stars, and a few \ac{NSBH} mergers~\cite{collaboration_gwtc-40_2026-1, collaboration_gwtc-40_2025-2, collaboration_gwtc-40_2025-1}\footnote{The exact number of sources depends on the threshold used to distinguish significant events, with recent catalogs typically using \ac{FAR}$< 1/~\mathrm{yr}$.}. Astrophysical black holes are characterized by only two parameters\footnote{\textcolor{black}{While black holes may also be charged, astrophysical black holes are expected to be effectively neutral due to interactions with surrounding plasma~\cite[e.g.,][]{Blandford:1977ds}.}}---their masses and spins---which leave imprints on the gravitational waveform of each compact-object merger event. Thus, analyses of the \ac{GW} strain data recorded in interferometric detectors, \textcolor{black}{which correspond to physical perturbations on the order of $\mathcal{O}(10^{-20})~\mathrm{m}$~\cite{Capote:2024rmo, TheLIGOScientific:2014jea, TheVirgo:2014hva, Aso:2013eba, Somiya:2011np, KAGRA:2020tym}}, allow us to constrain the parameters characterizing each binary and the population as a whole. These observations have been used for quantitative tests of general relativity, to constrain the cosmic expansion history, search for signatures of gravitational lensing, astrophysically calibrate our detectors, and learn about the astrophysical processes shaping the formation and evolution of compact-object binaries~\cite[e.g.,][]{LIGOScientific:2025pvj, collaboration_gwtc-40_2026, collaboration_gwtc-40_2026-2, collaboration_gwtc-40_2025, collaboration_gw240925_2026}\footnote{In the late stages of the preparation of this manuscript, the \ac{LVK} collaboration released their fifth catalog, GWTC-5, including over 100 new \ac{BBH} events~\cite{LIGOScientific:2026sit, LIGOScientific:2026ifv, LIGOScientific:2026wfs, LIGOScientific:2026ctl, LIGOScientific:2026uyd}. \textcolor{black}{This review focuses on the population of \ac{GW} black holes reported in GWTC-4 and exceptional events from later in \ac{O4}, but the general spin trends discussed here are robust in GWTC-5 data.}}.

The main theoretical \ac{BBH} formation channels are predicted to leave distinct imprints on their population-level spin distribution. However, the spins of individual \ac{GW} events are generally measured with large uncertainties, complicating the prospects of using spin as an astrophysical tracer. In this review, we provide an overview of gravitational-wave measurements of black hole spins, from astrophysical and dynamical predictions to individual-event and population-level measurements. We begin with an overview of the predictions of the main theoretical formation channels for black hole spin magnitudes and orientations in Section~\ref{sec:astro}. In Section~\ref{sec:measurements}, we describe the impact of spin on the gravitational waveform and characterize our ability to constrain spins for individual events, highlighting a few events with exceptional spin measurements. Population-level analyses are summarized in Section~\ref{sec:population}, including spin magnitudes, orientations, and correlations with other parameters. Finally, we conclude in Section~\ref{sec:conclusions} by highlighting some of the main open questions and future prospects for measuring spins with gravitational waves.

\section{Spin as a tracer of binary formation and evolution}
\label{sec:astro}
The formation channels of binary black holes fall broadly into two categories: isolated binary evolution and dynamical formation~\cite[see][for recent reviews]{Mandel:2018hfr, mapelli_formation_2021}. The canonical picture of isolated binary evolution in the field involves a common envelope phase following the formation of the first black hole that shrinks the orbit by several orders of magnitude via dynamical friction if the common envelope is successfully ejected~\cite[e.g.,][]{1976ApJ...207..574S, Postnov:2014tza, Tutukov:1993bse, Kalogera:2006uj, Ivanova:2012vx, Belczynski:2016obo}. Alternatively, \ac{SMT} may proceed via Roche-lobe overflow depending on the binary mass ratio, shrinking the orbit via angular momentum conservation~\cite{Inayoshi:2017mrs, Pavlovskii:2016edh, vandenHeuvel:2017pwp, Neijssel:2019irh}. Finally, binaries born in tight orbits in low-metallicity environments may avoid stellar mergers via chemically homogeneous evolution, leaving behind a \ac{BBH} system~\cite{Mandel:2015qlu, deMink:2016vkw, Marchant:2016wow}.

On the other hand, binaries may form via dynamical interactions including exchanges, direct collisions, and tidal captures that occur in dense environments like globular clusters or nuclear star clusters~\cite{PortegiesZwart:1999nm, Sigurdsson:1993zrm, Miller:2008yw, PortegiesZwart:2004ggg}. The binary is ``hardened'' by any strong gravitational encounter that drains orbital energy, shrinking the orbit. Around \ac{AGN}, gas torques can eventually align the orbits of stellar-mass black holes in the nuclear star cluster with the disk, enabling binary formation via migration and orbital hardening via angular momentum exchange with the surrounding gas~\cite[e.g.,][]{McKernan:2012rf, Bellovary:2015ifg, Bartos:2016dgn, Stone:2016wzz, Secunda:2018kar, tagawa_formation_2020}.

\begin{figure}
\includegraphics[width=\textwidth]{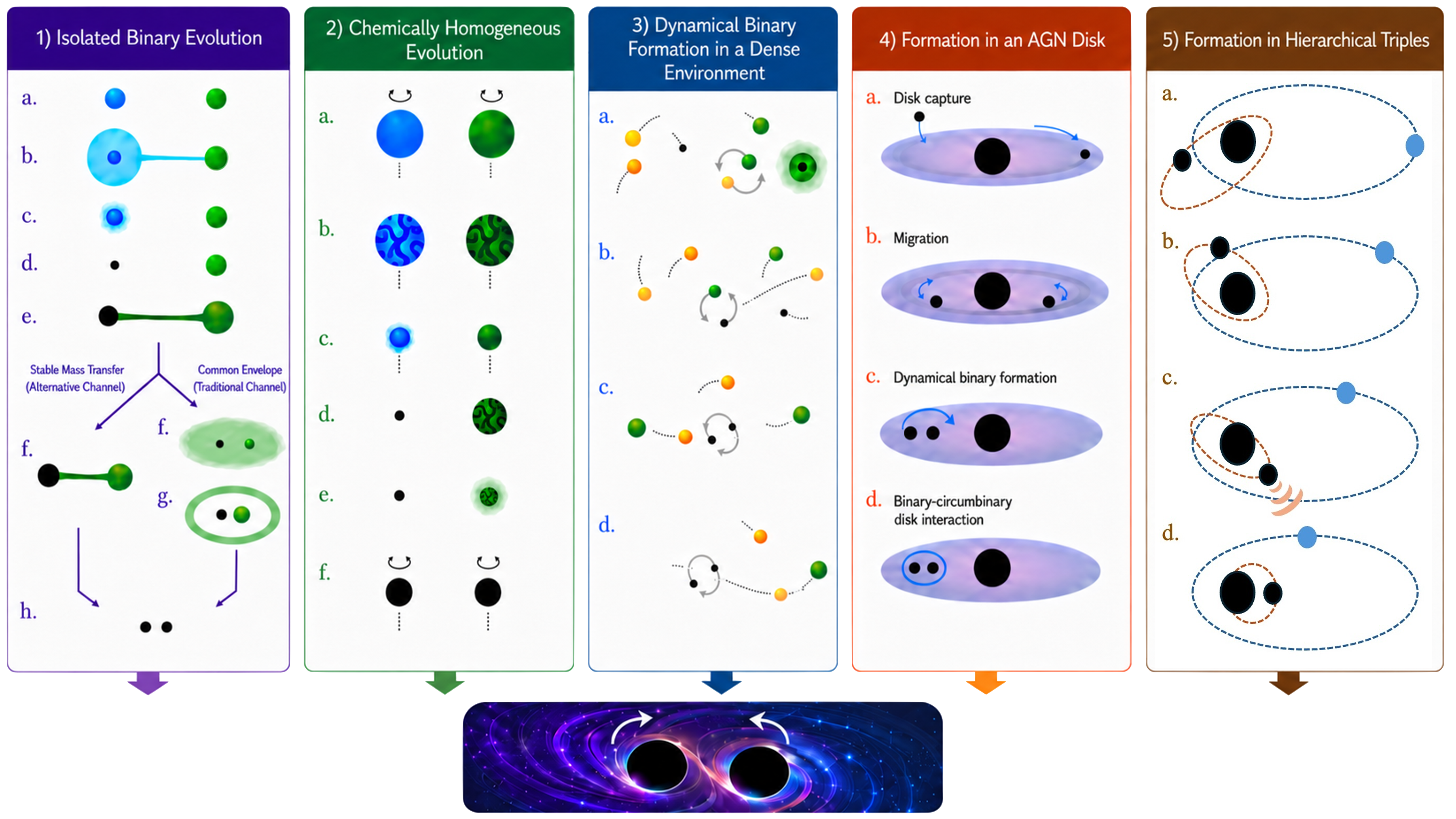}
\caption{\textcolor{black}{Schematic summarizing the formation of binary black holes via several leading formation channels: isolated binary evolution including either stable mass transfer or a common envelope phase, chemically homogeneous evolution, dynamical formation in clusters, formation in the disks of \acf{AGN}, and~hierarchical~triples. Panels 1-3 are adapted and reprinted from Ref.~\cite{Mandel:2018hfr} with permission from Elsevier.} \label{fig:formation_channels}}
\end{figure}

Finally, the formation of \acp{BBH} from stellar triples or higher-multiplicity systems incorporates elements of both isolated and dynamical formation~\cite[e.g.,][]{Silsbee:2016djf, liu_black_2018, Antonini:2017ash, antonini_precessional_2018, Stegmann:2021jen}, where Lidov--Kozai oscillations induced by the outer object excite significant eccentricity in the inner binary, leading to efficient orbital shrinking via the emission of bursts of gravitational radiation~\cite{vonZeipel, Lidov:1962wjn, Kozai:1962zz, Naoz:2016cjb}. Below, we describe the predictions for binary black hole spin magnitudes and angles for each of these pathways, which are summarized in Fig.~\ref{fig:formation_channels}.

\unskip

\subsection{Spin magnitudes}
The natal spin of a black hole depends primarily on the efficiency of angular momentum transport in its massive star progenitor. \textcolor{black}{Black hole spins are typically reported in terms of a dimensionless spin parameter, $\chi=|\vec{S}c/(Gm^{2})|$, where $m$ is the black hole mass and $\vec{S}$ its spin angular momentum, which reaches $\chi = 1$ for a maximally spinning black hole but can exceed $\chi \sim 10$ for rapidly rotating massive stars~\cite{Brott:2011ni, deMink:2012xx}}. Recent asteroseismic observations suggest that angular-momentum transport in low-mass stars is more efficient than theoretical models based on purely hydrodynamic processes had previously predicted~\cite[e.g.,][]{Beck:2011aa, Mosser:2012xc, Tayar:2013lra, Deheuvels:2014zha, fuller_angular_2014, Cantiello:2014uja, hermes_white_2017, gehan_core_2018}. Instead, the magnetohydrodynamical instability known as the (modified) Tayler–Spruit dynamo provides a viable mechanism to explain this enhanced efficiency~\cite{tayler_adiabatic_1973, spruit_dynamo_2002, fuller_slowing_2019, fuller_most_2019, ma_angular_2019}. In this process, magnetic torques convert any differential rotation inside the star to rigid-body rotation, allowing angular momentum to be transported from the contracting core to the outer layers of the star, which are eventually ejected as the star evolves. The end result is a slowly-spinning core that collapses to a slowly-spinning black hole, with natal spins on the order of $\chi \sim 0.01$ ($\chi \sim 0.1$) for the modified (classic) dynamo mechanism~\cite{fuller_most_2019, belczynski_evolutionary_2020, spera_compact_2022}. \textcolor{black}{The observation of maximally spinning black hole high-mass X-ray binaries challenges this picture, as spin-up is not expected to be effective enough over the short accretion timescales in these systems to achieve such high spins, albeit with potentially significant uncharacterized systematic measurement uncertainties~\cite[e.g.,][]{Reynolds:2020jwt}.}

Large natal spins are still possible in binary black holes where tidal spin-up is efficient. This can occur in binaries that form via \ac{CHE} from very massive, low-metallicity progenitors on tight orbits~\cite{Mandel:2015qlu, deMink:2016vkw, Marchant:2016wow}. The rotation rates of the stars become synchronized to the orbital rotation via tidal locking, ensuring efficient chemical mixing in their interiors, which prevents the development of the core-envelope structure required for efficient angular momentum transport. However, the pulsational pair-instability supernova process can facilitate angular momentum loss in the lowest-metallicity stars undergoing \ac{CHE}~\cite{Zevin:2022wrw, Bavera:2022mef}. Alternatively, tidal spin-up of the progenitor of the second-born black hole can occur in post-common-envelope binaries consisting of a black hole and a Wolf-Rayet star~\cite{Kushnir:2016zee, Hotokezaka:2017esv, zaldarriaga_expected_2018, Qin:2018vaa, Bavera:2020uch, Ma:2023nrf}. In both cases, the efficiency of tidal spin-up depends on the binary separation, which in turn depends on the metallicity and the progenitor mass via the efficiency of mass loss from stellar winds, which act to widen the binary. This implies that the black hole spin will be correlated with redshift (via the metallicity-dependent cosmic star formation history) and mass~\cite{Marchant:2016wow, Bavera:2022mef, Wang:2025swf}. Tidal-spin up is not expected to be efficient in binaries that form via \ac{SMT}, as these generally have larger orbital separations during the black hole--Wolf-Rayet phase~\cite{Zevin:2020gbd, Bavera:2022mef} (although see Ref.~\cite{olejak_unequal-mass_2024} for tidal spin-up predictions for \ac{SMT} binaries  with unequal masses). 

While black holes formed from stellar collapse in dynamical environments are expected to have small natal spins based on the efficient angular momentum transport arguments presented above, they can be spun up via accretion from stellar collisions. Recent work finds that up to $40\%$ of black holes in clusters may have spin $\chi \gtrsim 0.2$ due to previous stellar collisions, though the exact spin distribution is sensitive to the unknown accretion efficiency~\cite{Kiroglu:2024xpc}. A subpopulation of high-spin black holes is also expected for dynamically-formed binaries due to hierarchical mergers, where the remnant of a previous merger is retained in the cluster and goes on to merge again~\cite[see][for a review]{gerosa_hierarchical_2021}. These \ac{2G} black holes should have $\chi \approx 0.7$ based on conservation of angular momentum in the merger of two slowly-spinning, \textcolor{black}{approximately equal-mass} black holes~\cite{pretorius_evolution_2005, scheel_high-accuracy_2009, borchers_gravitational-wave_2025}. The probability of merger remnant retention and hence of hierarchical mergers depends sensitively on the escape speed of the environment, as the final black hole produced in a merger receives a recoil kick with velocity $v_{k}\sim \mathcal{O}(100)~\mathrm{km/s}$, the magnitude of which depends on the spins and mass ratio of the component black holes~\cite{Gonzalez:2006md, Campanelli:2007ew, Schnittman:2007sn, Lousto:2012su, Gerosa:2018qay}. For globular clusters with typical escape velocities $\sim 10-100~\mathrm{km/s}$, around $\sim 10\%$ of mergers may include a \ac{2G} black hole~\cite{Rodriguez:2019huv}. Mergers including higher-generation black holes are rare, as \ac{2G} black holes are rapidly spinning, which enhances the recoil kick velocity and reduces the retention probability~\cite{torniamenti_hierarchical_2024, Morawski:2018kfs, Antonini:2018auk, Zevin:2022bfa}. \textcolor{black}{In fact, recoil kick velocities can reach $v_k \sim 5000~\mathrm{km/s}$ for some spin configurations, which would lead to ejection from any host galaxy~\cite{Campanelli:2007cga, Gonzalez:2007hi, Lousto:2011kp, Merritt:2004xa}.} Meanwhile, mergers including a hierarchical component are predicted to be an order of magnitude more likely in nuclear star clusters where escape velocities can exceed $100~\mathrm{km/s}$~\cite{li_constraining_2022, sedda_fingerprints_2020, mapelli_mass_2021, Mapelli:2021syv} and in AGN disks, \textcolor{black}{where nearly all merger remnants are retained due to the deep potential well of the central supermassive black hole and the effect of gas damping~\cite{tagawa_formation_2020}. Up to $\sim 25 - 50\%$ of mergers in \ac{AGN} disks can include a hierarchical component~\cite{tagawa_formation_2020, Li:2022gul, Li:2022mck}}, and the rate of mergers including at least a third-generation black hole can exceed the rate of first-generation mergers later in the \ac{AGN} lifetime due to the efficiency of repeated mergers in migration traps~\cite{Secunda:2018kar, mckernan_mcfacts_2025, delfavero_mcfacts_2025}.

\subsection{Spin orientations}
\label{sec:astro_angles}
The spins of \acp{BBH} formed through isolated binary evolution are expected to be aligned to the orbital angular momentum due to the effects of tides in the progenitor stars, which work to align their spin axes to the orbit~\cite[e.g.,][]{Tutukov:1993bse, Kalogera:1999tq, Grandclement:2003ck, Postnov:2014tza, Belczynski:2016obo, Mandel:2015qlu, Marchant:2016wow, Rodriguez:2016vmx, Stevenson:2017tfq}. Moderate misalignments can be imparted by supernova natal kicks, in which an asymmetric supernova explosion tilts the orbital plane of the binary, misaligning the black hole spin axis~\cite{Kalogera:1996rm, Kalogera:1999tq}. The magnitudes and directions of black hole natal kicks are subject to significant uncertainties; theoretical modeling suggests that the magnitudes should be suppressed by fallback onto the black hole~\cite{Fragos:2010tm, Dominik:2012kk, Fryer:2011cx}, while some black hole X-ray binary observations are consistent with large $\sim 100~\mathrm{km/s}$ kick magnitudes~\cite[e.g.,][]{Willems:2004kk, Fragos:2008hg, Repetto:2015kra, Atri:2019fbx, Kimball:2022xbp}. Meanwhile, correlations between the proper motions and spin axes of some pulsars suggest that neutron star natal kicks may be preferentially oriented along their spin axes~\cite{Willems:2007jj, Johnston:2005ka, Noutsos:2013ce, Kaplan:2008qm}, but it is unknown if the same astrophysical processes leading to this correlation would apply for black holes. Binary population synthesis simulations generally assume that black hole natal kicks are isotropically distributed with magnitudes drawn from a Maxwellian distribution~\cite{Hobbs:2005yx, Hurley:2002rf, Giacobbo:2018etu, Breivik:2019lmt, COMPASTeam:2021tbl, fragos_posydon_2023} and predict black hole spin tilts $\lesssim 30^{\circ}$, with larger values of the velocity dispersion leading to larger tilts~\cite{gerosa_spin_2018}. \textcolor{black}{The degree of misalignment imparted by the kick depends primarily on the ratio of the kick velocity to the relative orbital velocity of the binary components and on the kick angle with respect to the orbital plane. For binaries that merge within a Hubble time, the required tight orbital separation at birth implies large orbital velocities and hence small misalignment angles. The kick velocity must be several times larger than the orbital velocity to achieve a large misalignment, but this increases the probability of binary disruption. As such,} tilts misaligned to the orbital angular momentum ($\theta > 90^{\circ}$) are difficult to explain with isolated binary evolution~\cite[e.g.,][]{Bavera:2020inc}. However, these arguments assume that the black hole inherits the spin direction of its progenitor star~\cite{Baibhav:2024rkn}, which may not be the case if instead the spin axis is tossed in a random direction at birth~\cite{Tauris:2022ggv}, as suggested for neutron stars by the large spin misalignment angles observed in some binary pulsars~\cite{Farr:2011gs}.

Meanwhile, the spins of black holes formed dynamically in clusters are expected to be isotropically distributed~\cite{Sigurdsson:1993zrm, Miller:2008yw, Zwart:2010kx, Benacquista:2011kv, Rodriguez:2016vmx}, as repeated dynamical interactions scramble the spin orientations. The hierarchical triple channel uniquely predicts an excess of black holes with in-plane spins~\cite{antonini_precessional_2018, rodriguez_triple_2018, liu_black_2018, yu_spin_2020, su_spin-orbit_2021} due to competing precession effects of the inner vs outer binaries.
The spins of the inner binary are initially aligned to its orbital angular momentum, $\vec{L}_{\mathrm{in}}$, as described above for isolated binaries, which must be nearly perpendicular to $\vec{L}_{\mathrm{out}}$ for mergers facilitated by Lidov-Kozai oscillations. At large separations, the spin evolution is dominated by the precession of $\vec{L}_{\mathrm{in}}$ around $\vec{L}_{\mathrm{out}}$, so the precession axis is aligned to $\vec{L}_{\mathrm{out}}$ and the angle between the spin vectors and the precession axis is $\theta_{p}\sim90^{\circ}$. At small separations, the precession axis gradually aligns to $\vec{L}_{\mathrm{in}}$, but $\theta_{p}$ is an adiabatic invariant, so the angle between the inner binary spin vectors and the precession axis, and hence $\vec{L}_{\mathrm{in}}$, is maintained to be perpendicular through the merger of the inner binary~\cite{stegmann_gravitational-wave_2026}.

Finally, the \ac{AGN} disk channel is expected to produce \acp{BBH} with spins either aligned or antialigned to the orbital angular momentum. This alignment occurs through the stellar-mass analog of the Bardeen-Petersen effect~\cite{Bardeen:1975zz}. First, any orbital inclination with respect to the disk is dissipated, so that the binary orbital angular momentum is either aligned or anti-aligned to the angular momentum of accretion disk~\cite{Lubow:2015, Moody:2019nes}. A circum-binary disk will form around the binary, with minidisks around each black hole, which exert gas torques to keep the black hole spin aligned to the angular momentum of the disks~\cite{King:2005mv}. While recent works suggest that the distribution of spin tilts for binaries formed in \ac{AGN} disks should favor aligned over anti-aligned binaries, the exact fraction is subject to considerable modeling uncertainties, affected by the unknown ratio of retrograde to prograde orbits~\cite{cook_mcfacts_2024}, the role of binary-single encounters~\cite{tagawa_spin_2020}, and the stability of retrograde configurations~\cite{dittmann_evolution_2024, mckernan_monte-carlo_2020}.

\section{Measuring spin with gravitational waves}
\label{sec:measurements}
\subsection{The effect of spin on the GW waveform}
\label{sec:spin_phys}
A quasi-circular compact-object binary is fully characterized by 17 free parameters, $\vec{\theta}$. Seven parameters describe the relative position and orientation of the binary with respect to the observer; \textcolor{black}{these include two parameters for the sky location of the source (right ascension and declination)}, the phase and time at coalescence, the luminosity distance and inclination of the orbital plane relative to the line of sight of the observer, and polarization angle. The remaining ten parameters are intrinsic to the binary, including the masses, tidal deformabilities (which are assumed to be zero for black holes), and spin vectors of each component, where the more massive component of the binary is called the primary ($m_1 > m_2$). The spins are typically parameterized in terms of the dimensionless spin magnitude, $\chi$, a tilt angle with respect to the orbital angular momentum, $\theta$, and two azimuthal angles, $\phi_{12}$ and $\phi_{JL}$. \textcolor{black}{Each of these parameters affects the gravitational waveform observed in the frequency band of ground-based \ac{GW} detectors, which can be described in the inspiral regime using the \ac{PN} approximation~\cite[e.g.,][]{blanchet_gravitational_2002}, providing an expansion of the General Relativistic equations of motion in orders of the velocity of the moving bodies. Binaries only spend the last few minutes to seconds before merger in this frequency band, compared to the $\mathrm{Myr-Gyr}$ delay times expected between formation and merger.}

\begin{figure}
\begin{adjustwidth}{-\extralength}{0cm}
\centering
\includegraphics[width=1.2\textwidth]{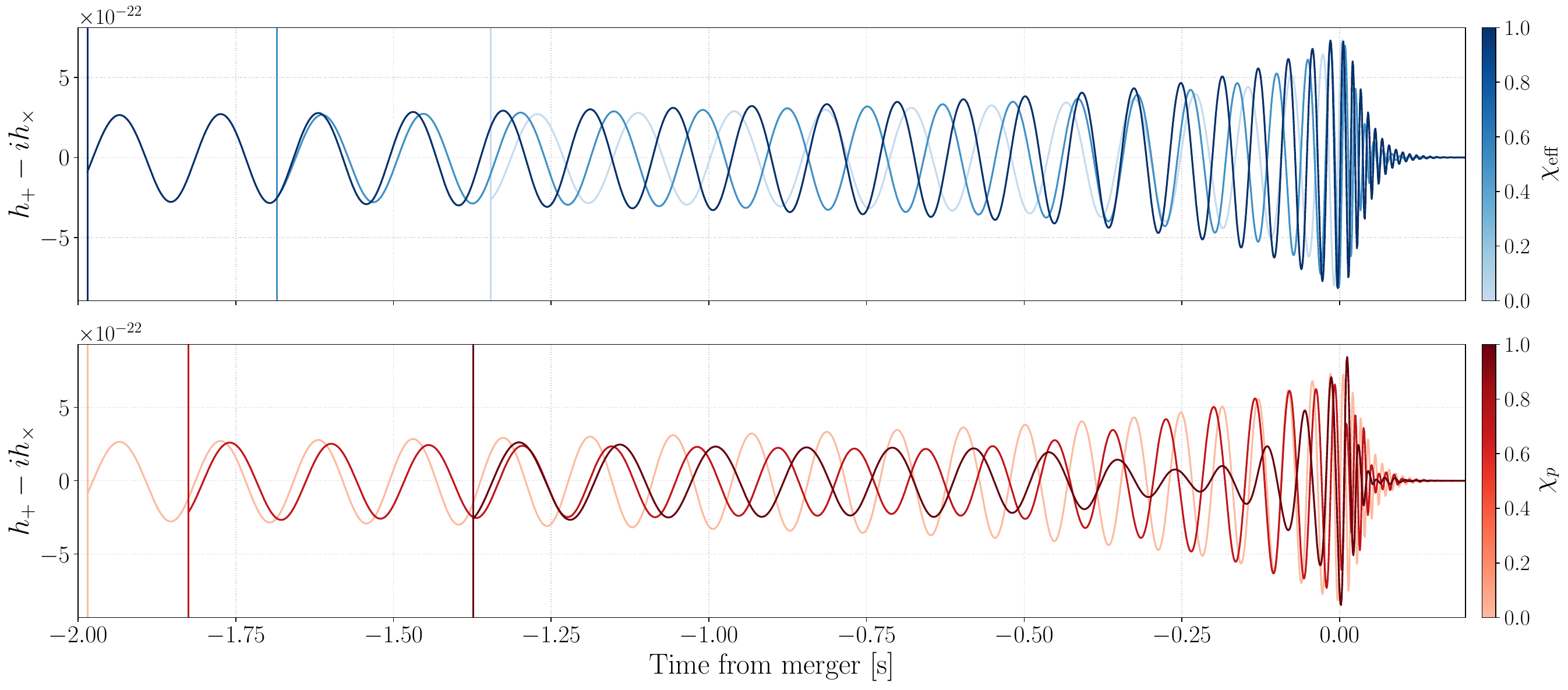}
\end{adjustwidth}
\caption{\textcolor{black}{Time-domain gravitational-wave strain for a simulated event with different values of \chieff (top) and $\chi_{p}$ (bottom), with all other binary parameters consistent with those inferred for the heavy \ac{BBH} merger GW190521. We use the SEOBNRv5PHM waveform model and a reference frequency of $11~\mathrm{Hz}$. The vertical lines show the start time of each waveform from a frequency of $6~\mathrm{Hz}$.} \label{fig:spin_waveform}}
\end{figure}

Spins have two primary effects on the evolution of a compact-object binary and its emitted gravitational radiation. If the spins are misaligned to the orbital angular momentum, $\vec{L}$, for most parameter configurations, the binary will undergo simple precession~\cite{apostolatos_spin-induced_1994}, \textcolor{black}{which imprints characteristic amplitude modulations on the waveform, as shown in the bottom panel of Fig.~\ref{fig:spin_waveform}}. In this case, the spin and orbital angular momentum vectors precess about the total angular momentum, $\vec{J}$, at a frequency that is small compared to the frequency of the \ac{GW} radiation. A \textcolor{black}{comparable-mass} binary that sweeps through the band of ground-based \ac{GW} detectors from $\sim 10 -500~\mathrm{Hz}$ will undergo $\mathcal{O}(10)$ precession cycles, with most of these precession cycles occurring at low frequencies~\cite{Cutler:1992tc}. 
The exact dynamics depend on the relative magnitude and orientation of $\vec{L}$ and $\vec{S}$; when $|\vec{S}|\leq |\vec{L}|$  or $\vec{S}$ and $\vec{L}$ are nearly aligned, the precession cone around $\vec{J}$ is narrower, leading to less extreme amplitude modulations in the waveform. The degree of amplitude modulation of the waveform also depends strongly on the inclination angle of the observer~\cite{apostolatos_spin-induced_1994, kidder_coalescing_1995}. This effect is enhanced for binaries with nearly edge-on inclinations, where the observer can see both sides of the orbital plane as it oscillates in the extreme case.

A special case of simple precession occurs when the binary configuration corresponds to a ``spin--orbit resonance''; the two component spin vectors and the orbital angular momentum vector align to form a 2D plane and jointly precess about $\vec{J}$~\cite{schnittman_spin-orbit_2004}. Rather than circulating freely, the azimuthal angle between the in-plane projections of the component spins is restricted to two specific values, $\phi_{12} = \pm \pi$ if $\theta_1 > \theta_2$ and $\phi_{12} = 0$ if $\theta_1 < \theta_2$~\cite{gerosa_resonant-plane_2013}. These resonant configurations are more likely to occur in isolated binaries, since the initial tilt angles must be small but nonzero. Binaries where the heavier black hole forms first (standard mass ratio) will end up with $\phi_{12} = \pm \pi$ since $\theta_1 > \theta_2$ on average as the first-formed black hole experiences two supernova kicks~\cite{gerosa_resonant-plane_2013, gerosa_spin_2018}. 

When spin-spin couplings become significant, simple precession is replaced with generic two-spin precession that includes nutation, in which the polar angle between $\vec{L}$ and $\vec{J}$ oscillates~\cite{kesden_effective_2015}. For some parameter configurations when $\vec{J}$ is small, a nutational resonance called ``transitional precession'' occurs, causing the total angular momentum vector to tumble in space, bringing $\vec{L}$ and $\vec{S}$ along with it~\cite{zhao_nutational_2017}. While nutation is expected to be prevalent among isolated binaries~\cite{steinle_signatures_2022}, the likelihood of observing the \ac{GW} emission from a binary undergoing transitional precession is low~\cite{apostolatos_spin-induced_1994}. In extreme cases, nutation can result in a complete flip of the black hole spin direction with respect to the orbital angular momentum~\cite{lousto_spin_2016, lousto_unstable_2016, gerosa_wide_2019}. A related dynamical instability occurs in ``up-down'' binaries, where the more massive black hole is aligned to the orbit and the secondary is anti-aligned~\cite{varma_up-down_2021, gerosa_precessional_2015}. Deviations from exact alignment in these ``up-down'' binaries lead to an instability that causes such systems to evolve towards the $\phi_{12}=0$ spin-orbit resonance~\cite{mould_endpoint_2020}.

Even for binaries with spins (anti)aligned to the orbital angular momentum, the spin contributes to the accumulated orbital phase~\cite{kidder_coalescing_1995}. \textcolor{black}{As illustrated in the top panel of Fig.~\ref{fig:spin_waveform}, the merger is delayed for systems with large prograde spins relative to a nonspinning binary with otherwise identical parameters due to the orbital hangup effect}, while for systems with large retrograde spins, the merger is hastened~\cite{campanelli_spinning-black-hole_2006}. The effect on the phasing is dominated by the contribution of spin--orbit couplings, which enter the waveform at 1.5-\ac{PN} order. Spin--spin couplings enter at 2-\ac{PN} and are generally weak unless the spin magnitudes are large~\cite{kidder_spin_1993, blanchet_gravitational_2002}. Both kinds of couplings affect both the phasing and the precessional dynamics described above. 

\textcolor{black}{The leading-order spin correction to the secular inspiral phasing depends only on the mass-weighted spin components aligned to the orbital angular momentum, which can be conveniently parameterized as}
\begin{align}
    \chieff = \frac{\chi_1\cos\theta_1 + q\chi_2\cos\theta_2}{1+q},
\end{align}
where $q = m_2/m_1 < 1$ is the binary mass ratio~\cite{Damour:2001, Ajith:2009bn, Ajith:2011ec, Santamaria:2010yb}. This parameter enters the \ac{GW} phase at 1.5-\ac{PN} order and is conserved through 2-\ac{PN} order when averaging over the orbital motion~\cite{racine_analysis_2008},
\begin{align}
    \psi_{1.5} &= (\pi\mathcal{M}f)^{-2/3}\psi,\\
    \psi &= \eta^{-3/5}\left[ \frac{(113-76\eta)}{128}\chi_{\mathrm{eff}} + \frac{76\delta\eta}{128}\chi_{a} - \frac{3\pi}{8}\right],
    \label{eq:spin_pn}
\end{align}
where $\chi_{a} = (\chi_1\cos\theta_1 - \chi_2\cos\theta_2)/2,\ \delta = (m_{1}-m_{2})/(m_{1}+m_{2})$, $\mathcal{M} = (m_1m_2)^{3/5}/(m_1 + m_2)^{1/5}$ is the chirp mass, and $\eta = q/(1+q)^{2}$ is the symmetric mass ratio. 
An analogous quantity has been defined to encapsulate the effects of precession~\cite{schmidt_towards_2015},
\begin{align}
   \chi_{p} &= \max{\left(\chi_{1}\sin{\theta_{1}}, \left(\frac{4q+3}{4+3q}\right)q\chi_{2}\sin{\theta_{2}}\right)}, \label{eq:chi_p}
\end{align}
though updated definitions that more robustly capture the effects of \ac{GW} radiation in higher-order multipoles and variations occurring on the
precessional timescale have been proposed~\cite{gerosa_generalized_2021, thomas_new_2021}. Alternatively, the measurability of precession can be captured via the precessing \ac{SNR}, by decomposing the waveform into a series of non-precessing harmonics that beat against each other to generate the amplitude and phase modulations imprinted by precession~\cite{Green:2020ptm, Fairhurst:2019srr, fairhurst_two-harmonic_2020}. 

\subsection{Measurability of spin in individual events}
The spins of compact-object binaries are measured using Bayesian inference, by comparing the \ac{GW} strain data to waveforms $h(\vec{\theta})$ from across the parameter space of interest. Stochastic sampling techniques are used to obtain samples from the posterior probability distribution on the binary parameters, where the likelihood of observing frequency-domain strain data $d_f$ given some choice of binary parameters $\vec{\theta}$ is given by~\cite[e.g.,][]{Romano:2016dpx, Veitch:2014wba}
\begin{align}
    \mathcal{L}(d_f | \vec{\theta}) \propto \exp\left[-\frac{2|d_f - h_f(\vec{\theta})|^2}{TS_{n,f}}\right],
\end{align}
where $T$ is the duration of the analyzed data segment and $S_{n,f}$ is the detector noise \ac{PSD}. \textcolor{black}{Mass priors are chosen based on the properties of each event identified by the matched-filter search pipelines, which currently cover the sub-solar to several-hundred solar mass regime, with mass ratios $q \lesssim 1/100$, though the exact details vary between search pipelines and observing runs~\cite[e.g.,][]{DalCanton:2017ala, Mukherjee:2018yra, Sakon:2022ibh, Allene:2025saz, LIGOScientific:2026ifv}. Spin priors for the analyses of individual events are typically assumed to be uniform in magnitude and isotropic in direction.}

\begin{figure}
\begin{adjustwidth}{-\extralength}{0cm}
\centering
\subfloat[\centering]{\includegraphics[width=0.6\textwidth]{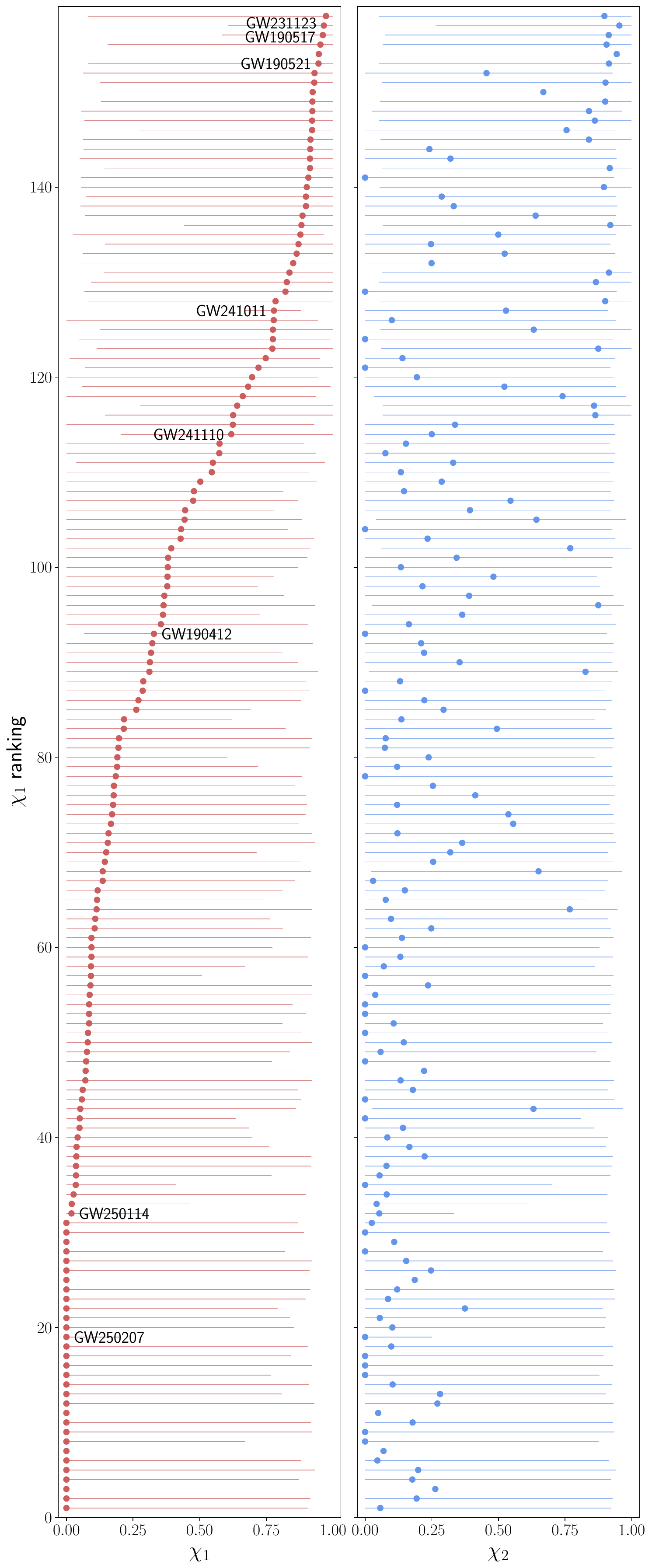}}
\subfloat[\centering]{\includegraphics[width=0.6\textwidth]{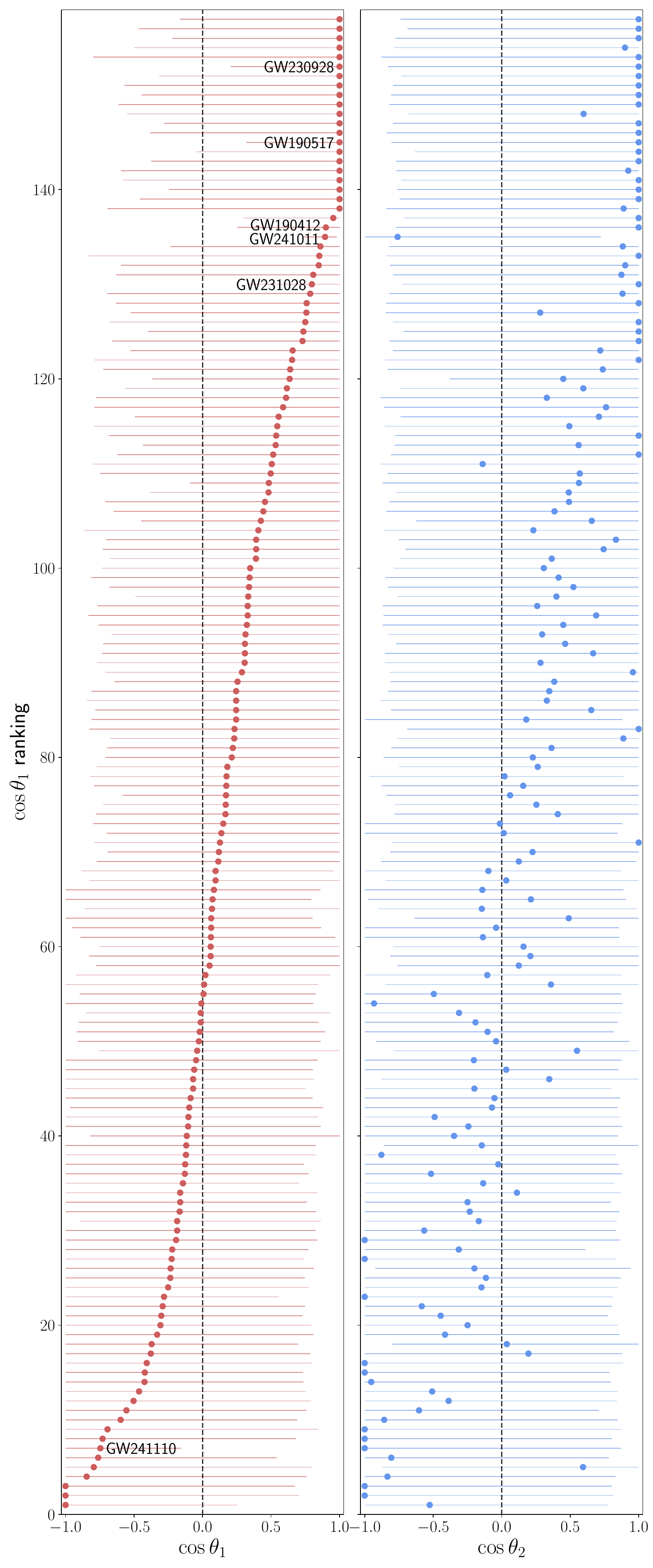}}
\end{adjustwidth}
\caption{\textcolor{black}{Maximum posterior value and highest-posterior density 90\% credible interval for the (\textbf{a}) spin magnitudes and (\textbf{b}) tilts for all the events with $\mathrm{FAR} < 1/\mathrm{yr}$ from GWTC-4, and the five exceptional events from later in \ac{O4}. Events are ordered by the maximum posterior value of the primary for magnitudes and tilts independently.} \label{fig:component_spins}}
\end{figure}

\begin{figure}
\begin{adjustwidth}{-\extralength}{0cm}
\centering
\includegraphics[width=1.3\textwidth]{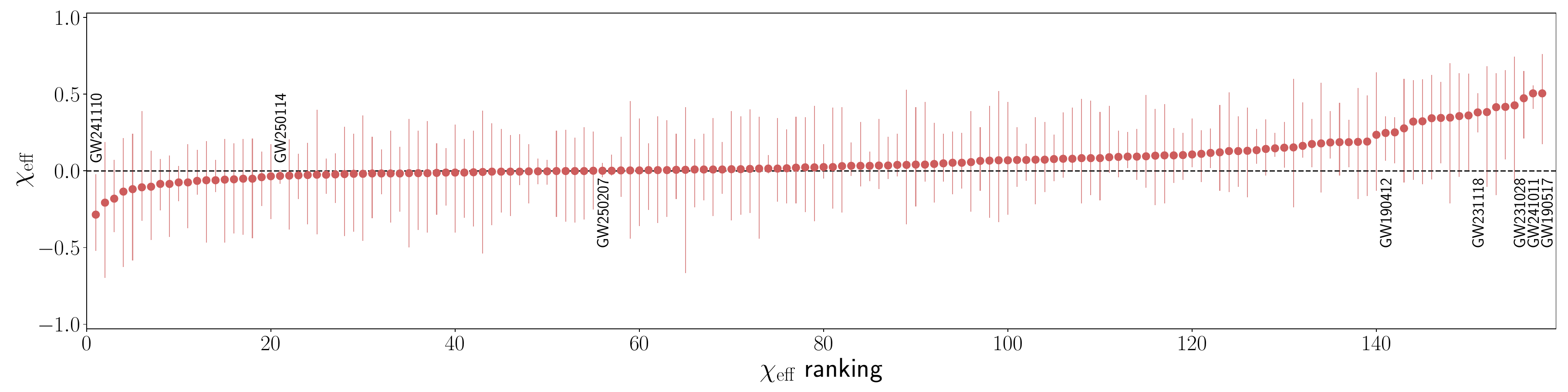}
\end{adjustwidth}
\caption{\chieff ordered by the maximum posterior value for all the events with $\mathrm{FAR} < 1/\mathrm{yr}$ from GWTC-4, and the five exceptional events from later in \ac{O4}.\label{fig:individual_chieff}. \textcolor{black}{The error bars span the highest-posterior density 90\% credible interval.}}
\end{figure}

In Figs.~\ref{fig:component_spins}-\ref{fig:individual_chieff}, we show the spin constraints on the individual component spin magnitudes, tilts, and \chieff for all the events in the fourth \ac{LVK} catalog (GWTC-4)~\cite{collaboration_gwtc-40_2025} and the five exceptional events from later in the \ac{O4}~\cite{collaboration_gw241011_2025, LIGOScientific:2025rsn, collaboration_gw240925_2026, LIGOScientific:2025rid}.
In general, the effective spin parameters \chieff and $\chi_{p}$ can be measured more accurately than the component spin magnitudes and angles~\cite{Purrer:2013ojf}, \textcolor{black}{for which current constraints are weak}. However, measurements of \chieff are strongly correlated with mass ratio along lines of constant $\psi$ (Eq.~\ref{eq:spin_pn}), particularly for low-mass, inspiral-dominated signals with non-precessing spins~\cite{baird_degeneracy_2013}, though a similar correlation with mass ratio has also been identified for heavy binaries~\cite{kang_mapping_2025}. Precession effects help to break this degeneracy~\cite{chatziioannou_spin-precession_2014}, leading to improvements in the measurements of both the spins and the extrinsic parameters~\cite{vanderSluys:2007st, lang_measuring_2011, pankow_astrophysical_2017}, since both sets of parameters dictate the extent to which amplitude and phase modulations are discernible in the waveform. Given that $|\vec{S}|$ dictates the precessional dynamics, spins are better measured in systems with large spin magnitudes~\cite{chatziioannou_measuring_2018}. 

Spins can also be measured more precisely for unequal-mass systems~\cite{pratten_measuring_2020}, as the term in Eq.~\ref{eq:spin_pn} proportional to $\delta$ vanishes for \textcolor{black}{exactly} equal-mass systems. Equal-mass systems are subject to a labeling degeneracy, where the spins of the more and less massive black holes cannot be distinguished. Instead, sorting the binary components by their spin magnitudes~\cite{Biscoveanu:2020are} or using machine learning methods to determine the optimal sorting scheme for each event~\cite{Gerosa:2024ojv} allows for more precise spin posteriors subject to fewer multimodalities. Including the \ac{GW} radiation in higher-order modes, which is enhanced for unequal-mass systems viewed at edge-on inclinations, improves both the accuracy and precision with which spins, and precession effects in particular, can be measured~\cite{cho_gravitational_2013, oshaughnessy_parameter_2014, oshaughnessy_parameter_2014-1, Green:2020ptm, krishnendu_interplay_2022}. Particularly for unequal-mass systems, two-spin effects become difficult to discern, and the spin of the secondary object is largely unconstrained~\cite{Purrer:2013ojf}. 

Early studies using waveform mismatch calculations suggested that configurations corresponding to different spin-orbit resonances may be distinguishable amongst themselves and from freely precessing systems~\citep{Gupta:2013mea, Gerosa:2014kta, Afle:2018slw}, but updated works using modern waveform models generally find that the spin tilt and azimuthal angles are difficult to measure~\cite{vitale_measuring_2014} with the precision needed to identify resonances, even with large signal-to-noise ratios~\cite{biscoveanu_measuring_2021}. While $\phi_{12}$ can be well-constrained in certain parts of the binary parameter space~\citep{Trifiro:2015zda, Johnson-McDaniel:2023oea}, the constraints are improved when measured at a reference frequency closer to merger, as the waveform is more sensitive to variations in these parameters in this regime~\citep{Varma:2021csh}.

\begin{figure}
\begin{adjustwidth}{-\extralength}{0cm}
\centering
\subfloat[\centering]{\includegraphics[width=6.0cm]{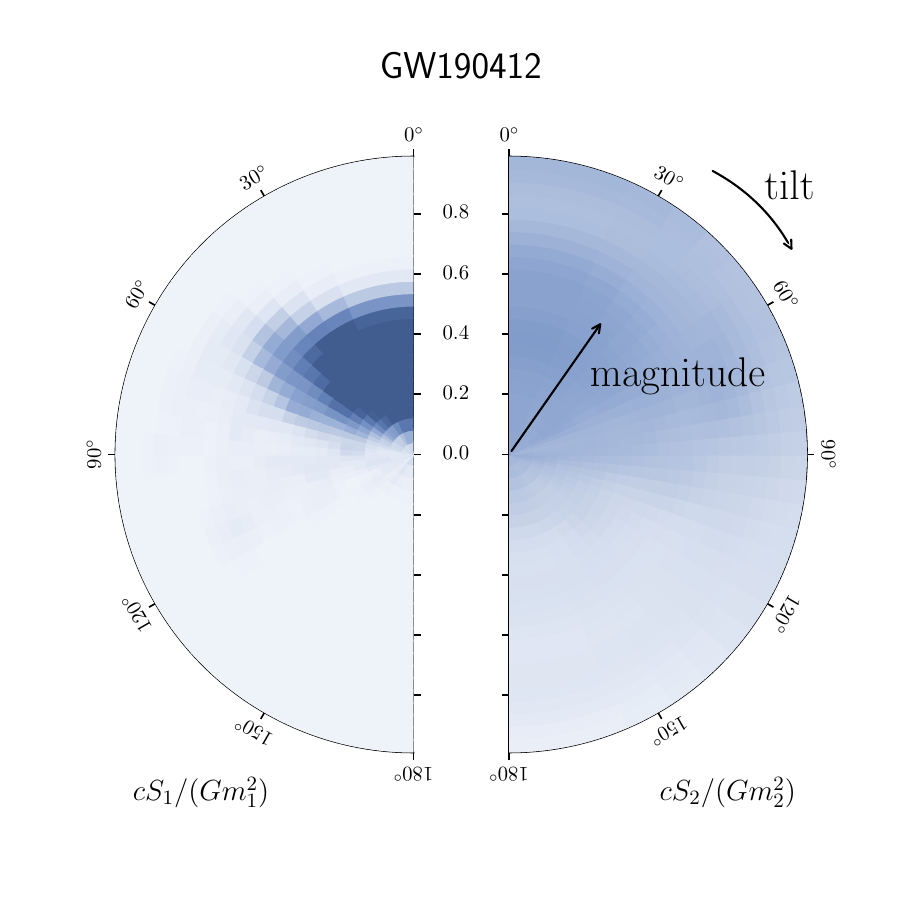}}
\subfloat[\centering]{\includegraphics[width=6.0cm]{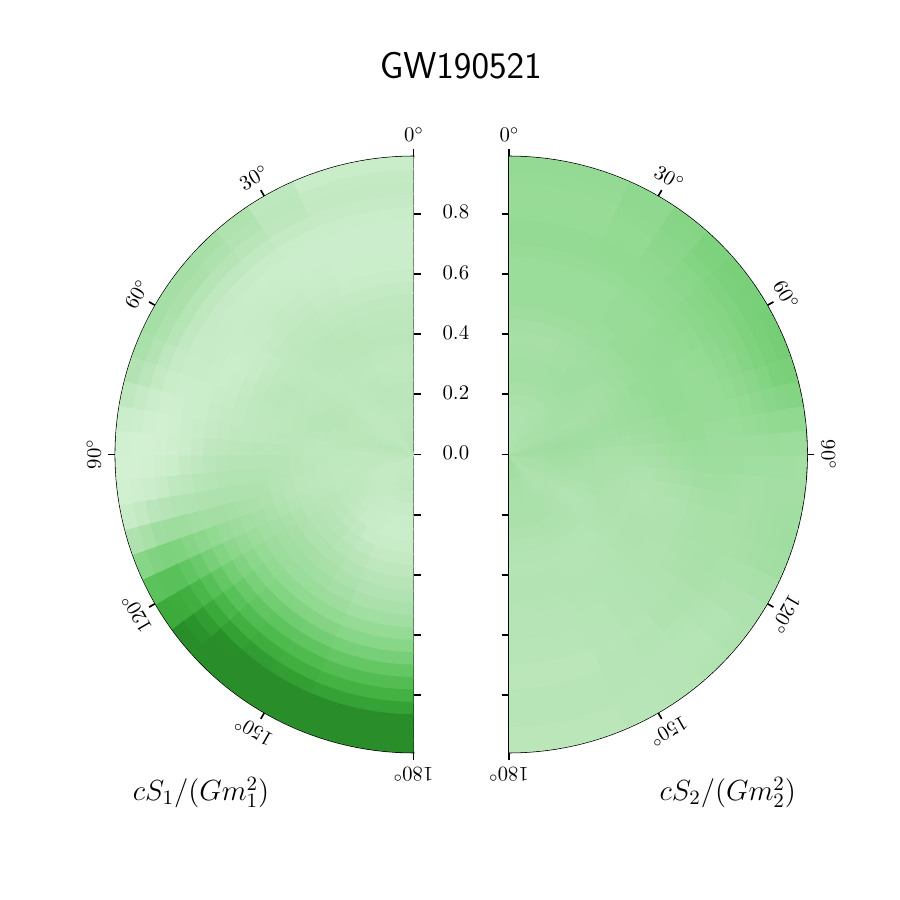}}
\hfill
\subfloat[\centering]{\includegraphics[width=6.0cm]{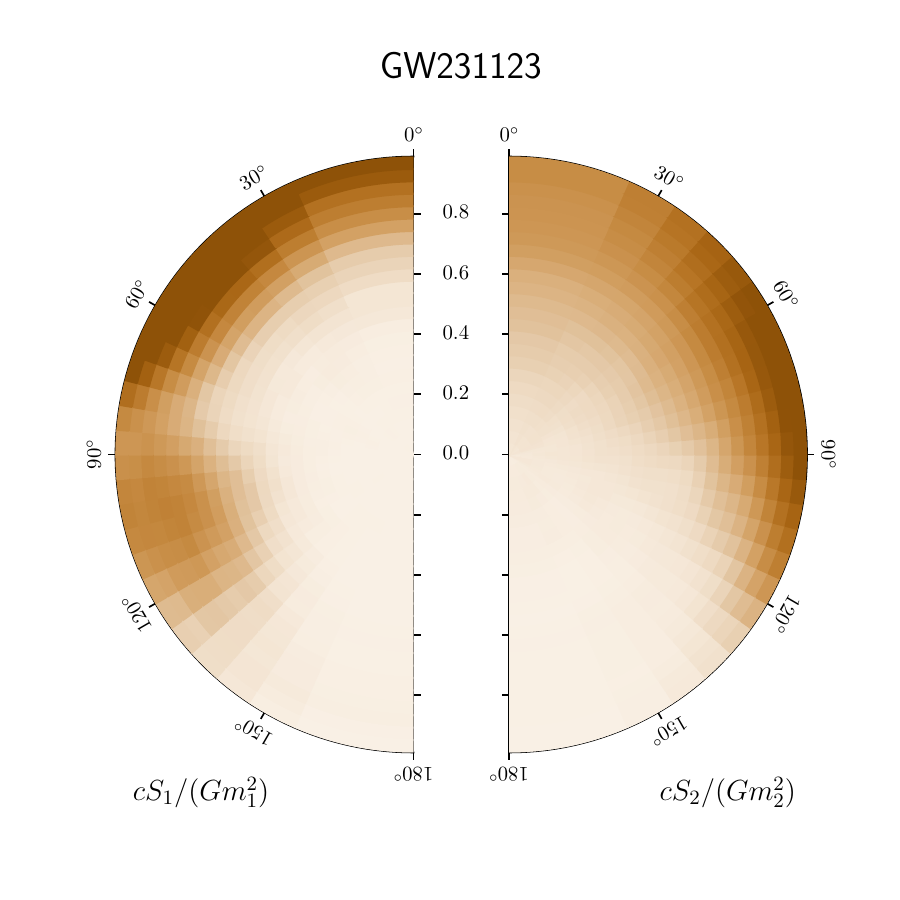}}
\subfloat[\centering]{\includegraphics[width=6.0cm]{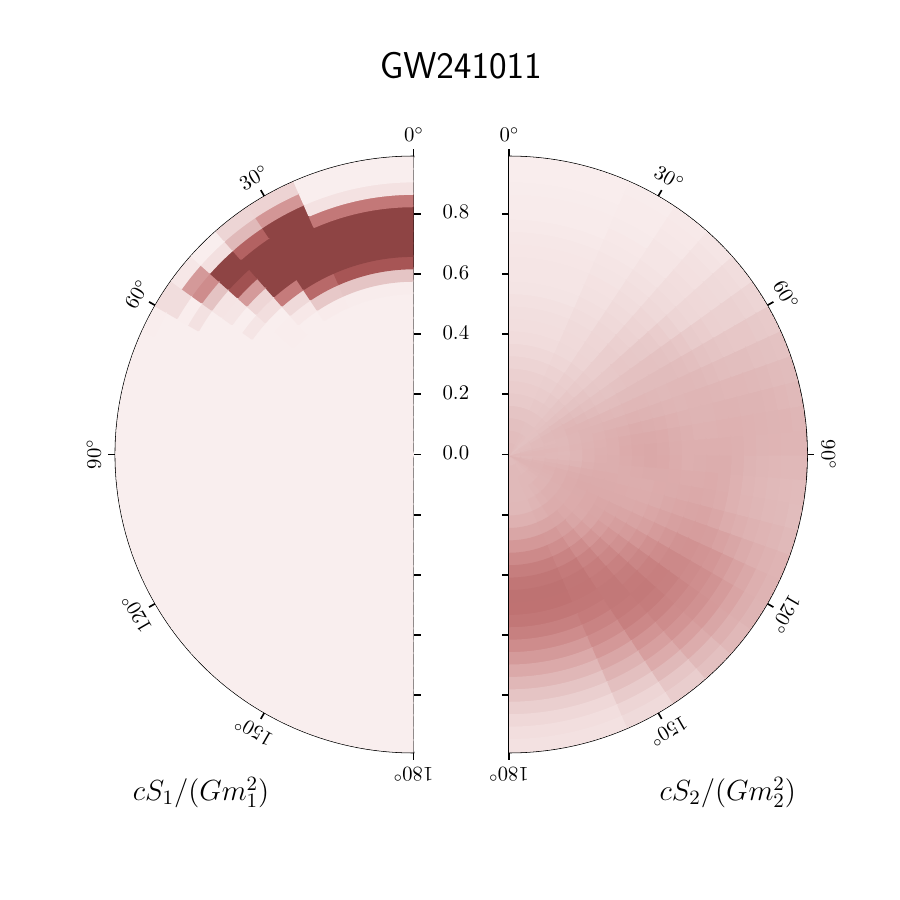}}
\subfloat[\centering]{\includegraphics[width=6.0cm]{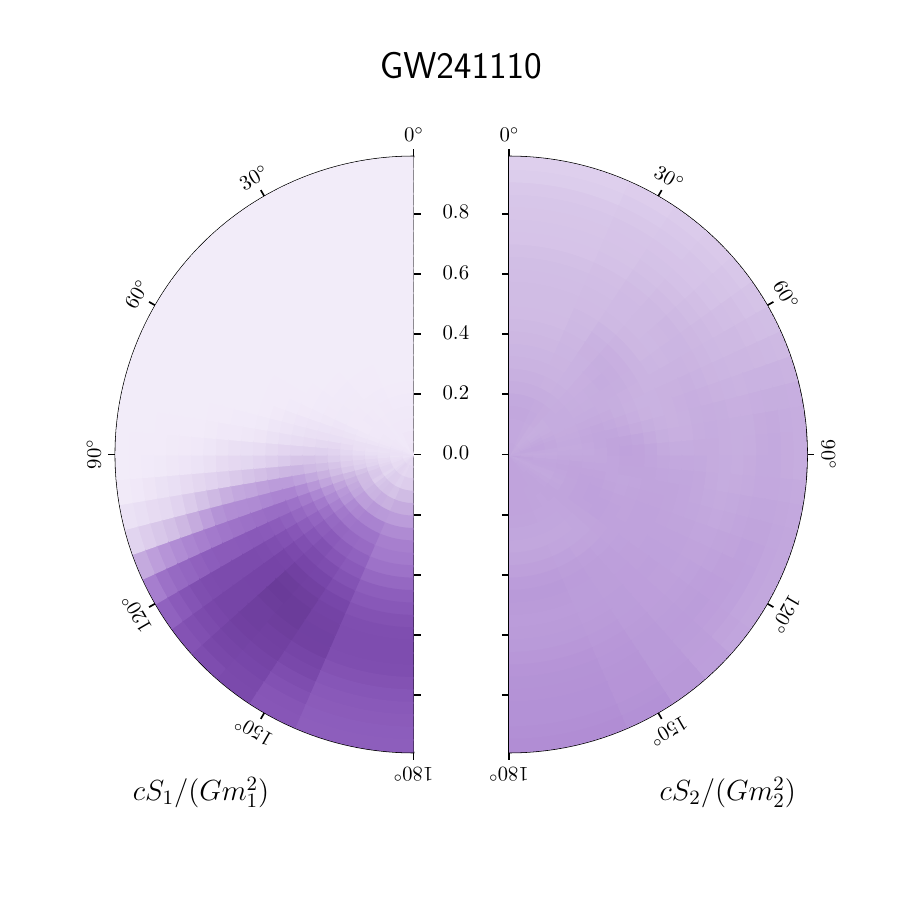}}
\end{adjustwidth}
\caption{Spin disk plots for exceptional \ac{GW} events with notable spin measurements: (\textbf{a}) GW190412, (\textbf{b}) GW190521, (\textbf{c}) GW231123, (\textbf{d}) GW241011, and (\textbf{e}) GW241110. The left (right) hemisphere of each panel shows the inferred spin of the primary (secondary) object. The radial direction indicates the spin magnitude and the angular direction the tilt angle with respect to the orbital angular momentum. \textcolor{black}{The shading conveys the posterior probability, with regions of higher probability indicated with a darker color in each plot.} \label{fig:spin_disk}}
\end{figure} 

A few individual \ac{BBH} events have noteworthy spin measurements, as shown in Fig.~\ref{fig:spin_disk}. %
GW190412 was the first event with a measurable unequal mass ratio ($q=\OThreeAsymMassRatioUncert$)\footnote{Reported quantities in this work are the maximum posterior value with highest-posterior density 90\% credible interval unless otherwise noted.} and contribution from higher-order modes~\cite{LIGOScientific:2020stg}. The primary object is confidently constrained to be spinning, with $\chi_1 = \OThreeAsymUncert$. More recently, the observation of the events GW241011\_233834 and GW241110\_124123 (henceforth GW241011 and GW241110, respectively) provided the strongest evidence of mergers formed hierarchically, as both of these events have unequal mass ratios and primary spin magnitudes characteristic of second-generation black holes, $q = \HierarchicalFirstMassRatioUncert,\ \chi_1 = \HierarchicalFirstUncert$ for GW241011 and $q = \HierarchicalSecondMassRatioUncert,\ \chi_1 =\HierarchicalSecondUncert$ for GW241110~\cite{collaboration_gw241011_2025}. Both primary spin vectors are misaligned to the orbital angular momentum, while likely pointing in opposite directions for the two events, another indication of dynamical formation. 

The observation of the event GW231123\_135430 (henceforth GW231123)~\cite{LIGOScientific:2025rsn} suggests the presence of a growing subpopulation of high-mass, high-spin events, as it joins GW190521 in this part of the parameter space~\cite{Abbott:2020tfl}. Both of these events broke the record for the most massive event at the time of their detection and both are consistent with the primary black hole being maximally spinning, with $M_{\mathrm{tot}} = \OFourHeavyMassUncert~M_{\odot},\ \chi_1 = \OFourHeavyUncert$ and $M_{\mathrm{tot}} = \OThreeHeavyMassUncert~M_{\odot},\ \chi_1 = \OThreeHeavyUncert$ for GW231123 and GW190521, respectively. The spin constraints on these high-mass events and signatures of precession are difficult to square with the intuition developed in the previous section which implies that spin information is predominantly imprinted in the inspiral, as only the last few inspiral cycles are detectable for such high-mass events, which are instead merger and ringdown-dominated in the band of the \ac{LVK} detectors. 

What drives the spin information for such massive events is one of the active, open questions in the field. Ongoing work suggests that uncertainties on the spin parameters do not necessarily increase monotonically with mass~\cite{vitale_parameter_2017, biscoveanu_measuring_2021}. For example, large aligned spins are better constrained for heavy systems than those without significant \ac{SNR} in the merger and ringdown. It is also easier to measure spin alignment than precession in heavy systems~\cite{vitale_parameter_2017, Purrer:2015nkh, biscoveanu_measuring_2021}. Noise fluctuations alone are not enough to explain the informative precession measurement in GW190521~\cite{biscoveanu_measuring_2021, xu_measurability_2023} or the high masses and spin magnitudes inferred for GW231123~\cite{bini_impact_2026}. For GW190521, the ability to obtain informative measurements of $\chi_{p}$ instead hinges on the relationship between the inspiral and merger/ringdown portions of the signal~\cite{miller_measuring_2025} and depends sensitively on the exact six-dimensional spin configuration~\cite{biscoveanu_measuring_2021}, while the choice of extrinsic parameters does not have as significant an effect~\cite{miller_measuring_2025}. Additionally, the short inspiral for such massive systems means that precession effects can be confused for eccentricity~\cite{Romero-Shaw:2020thy, Gayathri:2020coq, xu_measurability_2023, tibrewal_misinterpreting_2026}, since waveform models do not currently include both effects.

Spin inferences are also particularly sensitive to data quality issues. Non-Gaussian, transient noise artifacts known as ``glitches'' can bias spin measurements if they overlap with an astrophysical source~\cite{hourihane_accurate_2022, lecoeuche_coalescing_2026, udall_inferring_2026}. For example, the significance of the evidence for precession in GW200129\_065458 varies depending on the glitch mitigation technique~\cite{hannam_general-relativistic_2022, payne_curious_2022, macas_revisiting_2024}, and the interpretation of the misaligned spin in GW191109\_010717 also depends on the assumed glitch model~\cite{udall_anti-aligned_2025}. The inference of spin in short, high-mass systems like GW231123~\cite{ray_gw231123_2025} is particularly sensitive to data quality issues, especially since common glitch types \textcolor{black}{with similar durations and time-frequency morphologies} can masquerade as massive, high-spin \acp{BBH} when fit with an astrophysical source model~\cite{ashton_parameterised_2022}.

\section{The population of black hole spins}
\label{sec:population}
Once posteriors on the binary parameters $\vec{\theta}$ of each individual \ac{BBH} event are available, population-level analyses typically seek to constrain a set of hyper-parameters, $\vec{\Lambda}$, common to all events that describe the population-level distributions from which $\vec{\theta}$ are drawn. Some binaries are easier to detect than others (like those with large positive \chieff~\cite{ng_gravitational-wave_2018}), so the observed population is a biased representation of the underlying astrophysical distribution. The likelihood of observing an ensemble of events with data $\{d\}$ given hyper-parameters $\vec{\Lambda}$ accounts for these selection effects~\cite{Thrane:2018qnx, Mandel_2019, Vitale:2020aaz}:
\begin{align}
    \mathcal{L}(\{d\} | \vec{\Lambda}) &= \frac{1}{\alpha(\vec{\Lambda})^{N}}\int \mathcal{L}(\{d\} | \vec{\theta})\pi(\vec{\theta} | \vec{\Lambda})d\vec{\theta},\\
    \alpha(\vec{\Lambda}) &= \int p_{\mathrm{det}}(\vec{\theta})\pi(\vec{\theta} | \vec{\Lambda})d\theta,
\end{align}
where $\pi(\vec{\theta} | \vec{\Lambda})$ is the population model. In the selection effects correction, $\alpha(\vec{\Lambda})$ represents the probability of detecting an event drawn from a population with hyper-parameters $\vec{\Lambda}$, and $p_{\mathrm{det}}(\vec{\theta})$ is the probability of detecting an event with parameters $\vec{\theta}$, which is typically calculated using a simulated observing campaign~\cite{Essick:2023toz, Essick:2025zed}. 

\begin{figure}
\begin{adjustwidth}{-\extralength}{0cm}
\centering
\includegraphics[width=1.2\textwidth]{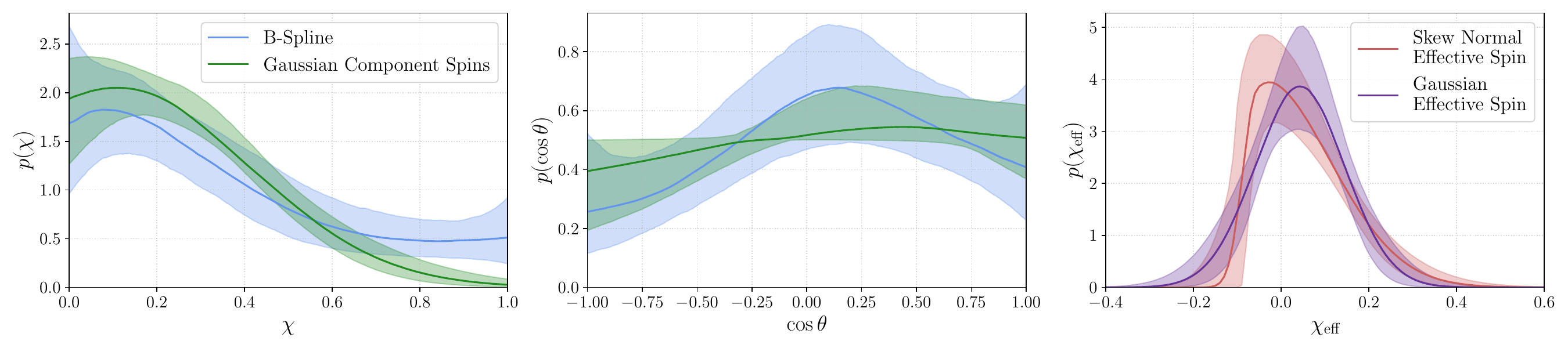}
\end{adjustwidth}
\caption{\textcolor{black}{Population-level spin distributions for the \acp{BBH} in GWTC-4 with $\mathrm{FAR} < 1/\mathrm{yr}$~\cite{LIGOScientific:2025pvj, ligo_scientific_collaboration_2025_16911563} . The component spin magnitude analyses do not distinguish between the primary and secondary spins in the binary. The shaded regions bound the 90\% credible intervals for each model and the solid lines represent the medians.} \textit{Left:} Inferred spin magnitude distribution using a strongly-modeled truncated Gaussian distribution (green) and a weakly-modeled B-spline (blue), \textit{Middle:} $\cos\theta$ distribution inferred using a phenomenological mixture model between isotropic and truncated Gaussian subpopulations (green) and B-splines (blue), \textit{Right:} $\chieff$ distribution inferred using skew normal (red) and truncated Gaussian (purple) population models\label{fig:spin_pop}}
\end{figure}

The population models themselves are usually characterized based on the strength of their assumptions, trading off interpretability for flexibility. \textit{Strongly-modeled} approaches include simple phenomenological distributions designed to capture broad features~\cite[e.g.,][]{Fishbach:2017zga, Talbot:2017yur, Miller:2020zox, banagiri_structure_2025}, \textcolor{black}{like the skew normal distribution used to fit the global peak and extended positive tail in the \chieff distribution shown in the right panel of Fig.~\ref{fig:spin_pop}}~\cite{banagiri_structure_2025}, and astrophysically-motivated models targeting specific effects, \textcolor{black}{like direct comparisons of the data against the outputs of population synthesis simulations to infer the common envelope efficiency or mixture fraction between specific formation channels}~\cite[e.g.,][]{Zevin:2020gbd, Wong:2020ise, Bouffanais:2021wcr, cheng_what_2023}. Meanwhile, \textit{weakly-modeled} or \textit{non-parametric} approaches use more generic distributions like splines~\cite{edelman_cover_2023, golomb_searching_2023, heinzel_probing_2024}, Gaussian processes~\cite{callister_parameter-free_2024} and mixture models~\cite{tiwari_vamana_2021}, and binned histograms~\cite{ray_non-parametric_2023, heinzel_pixelpop_2025} to let the data speak for itself. \textcolor{black}{These two approaches together provide a more complete picture of black hole population properties and can be compared against each other to identify model misspecification. For example, the increased probability at large spin magnitudes identified by the flexible spline model shown in the left panel of Fig.~\ref{fig:spin_pop} suggests that the strongly-modeled Gaussian distribution is insufficiently flexible to adequately capture this feature.}

Given the \textcolor{black}{distinct} spin predictions for different binary formation channels described above in Section~\ref{sec:astro}, population-level measurements of spins have long been regarded as a promising discriminator between \ac{BBH} formation channels. Early works using both Bayes factors~\cite{Vitale:2015tea} and hierarchical inference~\cite{Talbot:2017yur, Stevenson:2017dlk} typically predicted that 
the mixture fraction between channels could be measured using spin orientations (isotropic vs aligned) with as few as 100 events. If a single channel dominated and spin magnitudes were large, formation channels could be distinguished with as few as $\mathcal{O}(10)$ events~\cite{Stevenson:2017dlk, Farr:2017gtv, Wysocki:2018}. Similar optimistic predictions were made for the prospects of distinguishing hierarchical mergers~\cite{fishbach_are_2017} and using their unique features to measure mixture fractions between channels~\cite{Baibhav:2020xdf}. However, we are now in the regime of catalogs including \textcolor{black}{hundreds of} events, and the picture has proven more complicated that these early works imagined.

\unskip

\subsection{Spin magnitudes}
Strongly-modeled approaches using Beta distributions~\cite{Wysocki:2018, LIGOScientific:2018jsj, Abbott:2020gyp, KAGRA:2021duu} or Truncated Gaussians~\cite{LIGOScientific:2025pvj} have found consistent evidence for \textcolor{black}{an excess of} small spin magnitudes \textcolor{black}{among all black holes in compact-object binaries} since the first catalog of \ac{LVK} observations (GWTC-1), and this has been reinforced with non-parametric analyses of more recent catalogs~\cite{edelman_cover_2023, golomb_searching_2023, callister_parameter-free_2024} (see the left panel of Fig~\ref{fig:spin_pop}). A complementary picture emerges from analyses that fit the population-level distribution of \chieff instead of $\chi_{1/2}$ directly, where the global distribution of effective spin peaks at small but positive values, $\langle \chieff\rangle \sim 0.05$~\cite{Roulet:2018jbe, roulet_distribution_2021, Miller:2020zox, KAGRA:2021duu, LIGOScientific:2025pvj}. One of the open questions in the field is just how small these spin magnitudes are, and if there is evidence for a subpopulation of nonspinning black holes, which has been explored using both the spin magnitude~\cite{galaudage_building_2021, tong_population_2022, kimball_black_2020, kimball_evidence_2021, mould_which_2022, szemraj_disentangling_2026} and \chieff distributions~\cite{roulet_distribution_2021, callister_no_2022}. These analyses generally find that up to 80\% of the \ac{BBH} population can be nonspinning but cannot rule out the possibility that the entire population consists of binaries with two spinning black holes.

Most analyses looking at \ac{BBH} spin magnitudes assume that the spins of the primary and secondary black hole in each binary are independently and identically distributed from the same underlying astrophysical distribution. Under this assumption, the population-level distributions of the spin-sorted spin magnitudes suggest that the more rapidly spinning black hole in each binary (labeled $A$ rather than $1$ for the more massive) has typical spin $\chi_{A}\sim 0.4$, while the more slowly spinning black hole is consistent with $\chi_{B}\sim 0$~\cite{KAGRA:2021duu, szemraj_disentangling_2026}. However, this assumption has been relaxed in recently analyses to look for differences across the distributions of $\chi_{1/2}$, including those that may be imprinted by singly-spinning binaries as expected for tidal spin-up in systems formed via common envelope evolution. These works generally find modest evidence for two spin subpopulations, one with spin magnitudes around $\chi\sim0.1$ and a subdominant population consisting of $\sim 20\%$ of binaries with $\chi \sim 0.8$~\cite{adamcewicz_both_2025, hussain_hints_2024}. While conclusions about the secondary spin distribution vary based on the model assumptions and dataset~\cite{adamcewicz_no_2024, adamcewicz_both_2025, hussain_hints_2024}, determining which of the $\chi_{1/2}$ distributions favors higher spins encodes key information about formation channels, like the contribution of hierarchical mergers where $\chi_{1} > \chi_{2}$ is expected (see Section~\ref{sec:pop_correlations}), and the prevalence of mass ratio reversal in isolated binaries, which would imply that the more massive black hole formed second from the initially less massive star and was tidally spun up~\cite{mould_which_2022}.

Despite the small population of \ac{NSBH} mergers, early analyses suggest that the spins of black holes paired with neutron stars may be smaller than those paired with black holes~\cite{biscoveanu_population_2023, zhu_population_2022, LIGOScientific:2024elc}. Since most \ac{NSBH} systems are expected to form via isolated binary evolution~\cite{Mandel:2021smh}, the black hole \textcolor{black}{is only expected to} acquire significant spin if it forms second under mass ratio reversal, which may happen in up to $\sim 20\%$ of such systems~\cite{Broekgaarden:2021iew, chattopadhyay_modelling_2022}. A larger population will be required to determine if these early indications of small spins are robust and to begin probing \ac{NSBH} spin orientations.

\subsection{Spin orientations}
Spin tilts have typically been fit with a mixture model between aligned and isotropic components~\cite{Talbot:2017yur}, motivated by the predictions for isolated vs dynamical binaries discussed in Section~\ref{sec:astro_angles}. Recent works that relax the assumption of alignment in phenomenological models (as shown in the middle panel of Fig.~\ref{fig:spin_pop}) have found no evidence for a preferentially aligned subpopulation~\cite{vitale_spin_2022}, with a low-significance peak near in-plane spins instead emerging across both strongly and weakly-modeled analyses~\cite{edelman_cover_2023, callister_parameter-free_2024, golomb_searching_2023, stegmann_gravitational-wave_2026, LIGOScientific:2025pvj}. This excess of in-plane spins can be explained if a significant fraction of \acp{BBH} form in hierarchical triples, but more data will be required to determine if this is a robust population feature, as evidence for spurious peaks can be inferred with catalogs of the current size~\cite{vitale_long_2025}, and its exact properties and significance are model-dependent~\cite{wolfe_no_2026}. Beyond the spin tilt angles, analyses of the azimuthal angles are beginning to reveal hints of spin-orbit resonances on the population level~\cite{varma_hints_2022, biscoveanu_probing_2025}. However, simulation studies suggest that accurate and precise recovery of spin angle distributions is difficult for populations consisting of a mixture of aligned and isotropic systems, and that care should be taken when interpreting the measured mixture fraction parameter, as its recovery is model- and population-dependent~\cite{miller_gravitational_2024, vitale_long_2025, biscoveanu_probing_2025}.

While \chieff and $\chi_{p}$ are lossy parameters, their population-level distributions also carry information about spin orientations and are measured more precisely. Regardless of the population model used, the inferred \chieff distribution favors more aligned ($\cos\theta >0$) than antialigned systems~\cite{LIGOScientific:2025pvj}. Whether this preference is driven by a peak in the distribution at positive \chieff values or by a long tail at positive values in a distribution peaked at $\chieff=0$ depends on the population model (see the right panel of Fig.~\ref{fig:spin_pop})~\cite{Miller:2020zox, roulet_distribution_2021, banagiri_structure_2025}. Meanwhile, the $\chi_{p}$ distribution peaks away from zero, but its inference is more sensitive to analysis settings governing convergence of Monte Carlo integrals~\cite{talbot_growing_2023, LIGOScientific:2025pvj, kobayashi_impact_2026}. Systems with significant spin misalignments ($\cos\theta < 0$) are found to represent $\sim 20-40\%$ of the population using analyses of both $\chieff$ and $\cos\theta$, with implications for the fraction of \acp{BBH} formed dynamically~\cite{LIGOScientific:2025pvj}.

\subsection{Spin correlations}
\label{sec:pop_correlations}
While the results presented above have assumed that the \ac{BBH} mass, spin, and redshift distributions are independent, the formation channels discussed in Section~\ref{sec:astro} naturally imply correlations between these parameters, and ignoring such correlations can lead to bias in the inferred marginal distributions~\cite{alvarez-lopez_nowhere_2025}. Two main approaches have been explored to search for such correlations, evidence for which has been identified since the third \ac{LVK} catalog~\cite{KAGRA:2021duu}. The first approach explicitly models the dependence of the hyper-parameters governing one distribution on the other binary parameters. For example, Refs.~\cite{safarzadeh_trend_2020, callister_who_2021, biscoveanu_binary_2022} all allow the mean and width of the \chieff distribution to evolve linearly with mass and/or redshift, while Ref.~\cite{heinzel_probing_2024} relaxes this assumption and fits the relationship with a spline. The \chieff distribution is found to broaden with redshift, which indicates that \acp{BBH} were born with larger spin in the cosmic past~\cite{biscoveanu_binary_2022, heinzel_probing_2024, LIGOScientific:2025pvj}. Such a trend could be explained by tidal spin-up, which is more efficient in close binaries that also merge faster at higher redshifts, by reduced efficiency of stellar winds carrying away angular momentum in \ac{BBH} progenitors at low metallicities and high redshifts, or by an increasing fraction of hierarchical mergers at large redshift~\cite{Bavera:2022mef, belczynski_evolutionary_2020, farah_steep_2026}. There is also evidence for a correlation between either the \chieff mean or width and the binary mass ratio~\cite{callister_who_2021, adamcewicz_unequal-mass_2022, adamcewicz_evidence_2023, LIGOScientific:2025pvj}, which could be explained by the \ac{AGN} channel, in which hierarchical mergers are more common, and the heavier black hole is more likely to have been spun-up and aligned due to accretion in the disk~\cite{mckernan_monte-carlo_2020, cook_mcfacts_2024}.

Another approach involves looking for subpopulations across the mass, spin, and redshift parameter space with distinct properties~\cite[e.g.,][]{godfrey_cosmic_2024, banagiri_evidence_2025, galaudage_compactness_2025, ray_astrophysical_2026, berti_inferring_2025, afroz_binary_2025, afroz_phase_2025, Cheng:2026bpc, Padhyegurjar:2026scg, Padhyegurjar:2026slt}. The GWTC-4 data are beginning to reveal signatures of three sub-populations: one peaked at $10~M_{\odot}$ with a broad range of mass ratios and small but aligned spins consistent with isolated binary evolution, another peaked around $30~M_{\odot}$ strongly favoring equal mass ratios and isotropically-distributed spins that could correspond to first-generation mergers in clusters~\cite{rodriguez_dynamical_2016, park_black_2017, antonini_coalescing_2023, sedda_isolated_2026}, and a smaller subpopulation including both high masses and spins whose properties are more uncertain which could represent the confluence of multiple subdominant formation channels including hierarchical mergers, \ac{CHE}, and formation in \ac{AGN} disks~\cite{li_revealing_2025, li_aligned_2025}. The low-mass subpopulation can be explained via the \ac{SMT} channel~\cite{van_son_no_2022} or as a consequence of ``compactness peaks'' in the distribution of progenitor masses that decrease explodability and hence increase the likelihood of black hole formation~\cite{burrows_physical_2024, schneider_bimodal_2023, willcox_good_2025, galaudage_compactness_2025, Galaudage:2026opk}. 

Analyses targeting evidence for hierarchical mergers have employed both approaches, including mixture models between generational subpopulations~\cite{kimball_black_2020, kimball_evidence_2021, doctor_black_2020, li_constraining_2022} and looking for correlations between the spin distribution and the primary mass~\cite{antonini_star_2024, antonini_gravitational_2025, tong_subpopulation_2025} or mass ratio~\cite{plunkett_signatures_2026, vijaykumar_maximum_2026}. Current data hint at the presence of two subpopulations of hierarchical mergers, one at low primary masses consistent with GW241011 and GW241110, and another above the pair instability supernova mass gap around $\sim 45~M_{\odot}$~\cite{Fowler:1964zz, Barkat:1967zz, 1967ApJ...148..803R, 1968Ap&SS...2...96F, Heger:2001cd, Woosley:2007qp, Woosley:2016hmi, Farmer:2019jed}, where black holes are not expected to form from direct stellar collapse~\cite{tong_evidence_2025, tong_subpopulation_2025, antonini_gravitational_2025, plunkett_signatures_2026}. These hierarchical subpopulations may have different mass ratio and redshift distributions from the rest of the population, potentially explaining the observed \chieff--mass ratio and \chieff--redshift correlations~\cite{farah_steep_2026, vijaykumar_maximum_2026}.

\section{Conclusions and open questions}
\label{sec:conclusions}
The first decade of gravitational-wave observations has provided a new window into the spins of stellar-mass black holes. Although expectations based on observations of black hole X-ray binaries were that black holes in merging compact-object binaries would have large spins~\cite[e.g.,][]{Reynolds:2020jwt}, the growing catalog of gravitational-wave sources has revealed that these objects generally have small spin magnitudes. This makes it more difficult to obtain precise measurements of the component spins both for individual events and on the population level. Despite this challenge, gravitational-wave spin constraints have already provided a bounty of astrophysical insights into the processes shaping the formation and evolution of compact-object binaries.

The small observed spin magnitudes in most binaries are consistent with the picture of efficient angular momentum transport in their stellar progenitors. The preference for positive effective spins suggests that isolated binary evolution dominates among the formation channels, as large misalignments required for negative \chieff values are difficult to explain with this channel. However, the evidence for misaligned tilts and negative \chieff values on the population level indicates that some binaries form dynamically, even if this channel is subdominant. This picture is corroborated by recent analyses looking for \ac{BBH} subpopulations and correlations in the mass--spin--redshift parameter space. Evidence for hierarchical mergers continues to grow, both via the detection of individual events and a subpopulation of \acp{BBH} with the signature spin of $\chi\approx 0.7$ predicted for the remnant of a previous merger. Meanwhile, tentative evidence for alternative formation channels is beginning to appear with low significance, like the correlation between \ac{BBH} mass ratio and effective spin that can be naturally explained by the \ac{AGN} channel, or the hint of an excess of in-plane spins that only the hierarchical triple channel predicts.

In spite of this significant progress in connecting gravitational-wave spin observations to theory, several open questions remain. One key question revolves around the measurability of spin in high-mass systems, and how precession effects in particular are encoded in the waveform over only a few inspiral cycles. The ability to measure spin tilts on a population level represents another significant open question, as this parameter has long been heralded as a promising way to distinguish formation channels, but its population-level measurements are prone to biases and spurious features, even as the population grows. Finally, the tension between the small spins inferred with gravitational waves and the large spins observed in X-ray binaries remains unexplained. Significant development will be required to achieve parity between the understanding of X-ray binary selection effects and spin measurement systematics compared to the gravitational-wave case, where such errors are currently subdominant to large statistical uncertainties.

The next decade of gravitational-wave astronomy will see the growth of the catalog of compact-object binaries by an order of magnitude~\cite{KAGRA:2013rdx}, \textcolor{black}{and the growth of the detector network to include LIGO-India}~\cite{Saleem:2021iwi}, before next-generation detectors come online to detect hundreds of \ac{BBH} mergers per day~\cite{Evans:2021gyd, Branchesi:2023mws}. This increase in detector sensitivity will facilitate improvements in the spin constraints for individual events~\cite[e.g.,][]{Vitale:2016icu, Knee:2021noc, Pieroni:2022bbh, Cho:2022awq}, which will be detected with increasing \ac{SNR}. The detection of the astrophysical stochastic gravitational-wave background of \ac{BBH} mergers before the era of next-generation detectors will also provide a glimpse into the spin properties of high-redshift sources that are individually undetectable~\cite{Smith:2020lkj}, allowing us to better constrain population-level correlations. These advances will provide a sharper view of the spins of black holes in merging compact-object binaries, revealing a concordant picture of their formation and evolution.
\vspace{6pt} 

\funding{S.B. is supported by NSF grant number PHY-2513246.}
\dataavailability{All data behind the figures and values quoted in this manuscript are publicly available on Zenodo~\cite{ligo_scientific_collaboration_and_virgo_2021_5546663, ligo_scientific_collaboration_and_virgo_2022_6513631, ligo_scientific_collaboration_and_virgo_2025_16053484, ligo_scientific_collaboration_virgo_coll_2025_16877102,ligo_scientific_collaboration_2025_16911563,ligo_scientific_collaboration_2026_18600070}.} 

\acknowledgments{S.B. thanks Michael Zevin for completing the internal \ac{LVK} document review. She also thanks the organizers of the ``Taking Spin Measurements for a Spin: Recent Progress on Black Hole Spin Measurements Across the Electromagnetic and Gravitational Spectra'' Workshop at Wake Forest University in September 2025 for triggering insightful discussions that led to the preparation
of this manuscript.
This material is based upon work supported by NSF's LIGO Laboratory which is a major facility fully funded by the National Science Foundation. LIGO was constructed by the California Institute of Technology and Massachusetts Institute of Technology with funding from the National Science Foundation and operates under cooperative agreement PHY-0757058. 
The author is grateful for computational resources provided by the Caltech LIGO Laboratory supported by NSF PHY-0757058 and PHY-0823459.
This paper carries LIGO document number LIGO-P2600283.
During the preparation of this manuscript, ChatGPT v5.5 was used for the purposes of synthesizing references and generating Fig.~\ref{fig:formation_channels}. The author has reviewed and edited the output and takes full responsibility for the content of this publication.}

\conflictsofinterest{The author declares no conflicts of interest.}

\abbreviations{Abbreviations}{
The following abbreviations are used in this manuscript:
\\

\noindent 
\begin{tabular}{@{}ll}
O4 & fourth observing run\\
BBH & binary black hole\\
LVK & LIGO-Virgo-KAGRA\\
PN & post-Newtonian\\
SNR & signal-to-noise ratio\\
GW & gravitational-wave\\
PSD & power spectral density\\
CHE & chemically homogeneous evolution\\
AGN & active galactic nuclei\\
SMT & stable mass transfer\\
2G & second-generation\\
NSBH & neutron star-black hole\\
FAR & false alarm rate
\end{tabular}
}

\begin{adjustwidth}{-\extralength}{0cm}

\reftitle{References}

\bibliography{spin_review.bib, spins.bib}

\PublishersNote{}
\end{adjustwidth}
\end{document}